\documentclass[conference]{IEEEtran}
\IEEEoverridecommandlockouts

\usepackage{cite}
\usepackage{amsmath,amssymb,amsfonts}
\usepackage{algorithmic}
\usepackage{graphicx}
\usepackage{textcomp}
\usepackage{subcaption}
\usepackage[capitalise]{cleveref}
\usepackage{booktabs}
\usepackage[table]{xcolor}
\usepackage{microtype}
\usepackage{url}
\usepackage{enumitem}
\usepackage{numprint}

\def\BibTeX{{\rm B\kern-.05em{\sc i\kern-.025em b}\kern-.08em
    T\kern-.1667em\lower.7ex\hbox{E}\kern-.125emX}}

 \title{Distillation and Pruning for Scalable Self-Supervised Representation-Based Speech Quality Assessment}

\author{
\IEEEauthorblockN{\hspace*{-17mm}Benjamin Stahl$^*$} 
\IEEEauthorblockA{\hspace*{-17mm}\textit{Institute of Electronic Music and Acoustics} \\
\hspace*{-17mm}\textit{University of Music and Performing Arts}, 
Graz, Austria \\
\hspace*{-17mm}stahl@iem.at}
\and
\IEEEauthorblockN{Hannes Gamper}
\IEEEauthorblockA{
\textit{Microsoft Research}\\
Redmond, WA, USA \\
hannes.gamper@microsoft.com}
}

\newcommand\blfootnote[1]{%
  \begingroup
  \renewcommand\thefootnote{}\footnote{#1}%
  \addtocounter{footnote}{-1}%
  \endgroup
}

\begin{document}
%
\maketitle

\begin{abstract}
\blfootnote{$^*$Work done while an intern at Microsoft Research.}
\blfootnote{$^\textrm{1}$\url{https://github.com/microsoft/Distill-MOS}}
In this paper, we investigate distillation and pruning methods to reduce model size for non-intrusive speech quality assessment based on self-supervised representations. Our experiments build on XLS-R-SQA, a speech quality assessment model using wav2vec 2.0 XLS-R embeddings. We retrain this model on a large compilation of mean opinion score datasets, encompassing over 100,000 labeled clips. For distillation, using this model as a teacher, we generate pseudo-labels on unlabeled degraded speech signals and train student models of varying sizes. For pruning, we use a data-driven strategy. While data-driven pruning performs better at larger model sizes, distillation on unlabeled data is more effective for smaller model sizes. Distillation can halve the gap between the baseline's correlation with ground-truth MOS labels and that of the XLS-R-based teacher model, while reducing model size by two orders of magnitude compared to the teacher model.
\end{abstract} 

\begin{IEEEkeywords}
speech processing, quality of experience
\end{IEEEkeywords}

\section{Introduction}\label{sec:intro}
Complementing time-consuming and costly subjective evaluations of speech processing and generation systems with computational methods is a key objective in speech processing. Deep learning-based methods and increasing amounts of quality-rating labeled data have driven advances in computational speech quality assessment. A notable advantage of deep learning in this context is its ability to assess quality without a reference (e.g., DNSMOS, NISQA \cite{avila2019,reddy2021_3,mittag2021}).

Recently, self-supervised representation learning (SSL) based speech and audio models have been employed as feature extractors for speech quality assessment. Tseng et al.~\cite{tseng2021} explored various SSL models for assessing voice conversion quality, finding SSL embeddings to outperform Mel-frequency cepstral coefficients or Mel spectrograms, with wav2vec\,2.0 \cite{baevski2020} embeddings achieving the highest correlation with subjective ratings. Zezario et al.\ (MOSA-Net-Cross-Domain \cite{zezario2023}) combined HuBERT \cite{hsu2021} embeddings with learnable filterbanks and short-time Fourier transform features to predict both full-reference objective metrics and subjective ratings without a reference signal. Saeki et al.\ (UTMOS \cite{saeki2022}) fine-tuned wav2vec\,2.0 embeddings to predict subjective ratings of generated speech, ranking first in the 2022 VoiceMOS challenge \cite{huang2022}. Tamm et al.\ (XLS-R-SQA \cite{tamm2023}) employed wav2vec\,2.0\,XLS-R \cite{babu2022} features found via grid search, with XLS-R-SQA ranking first in the ConferencingSpeech 2022 challenge \cite{yi2022}. Deshmukh et al.\ (PAM \cite{deshmukh2024}) have used CLAP \cite{elizalde2023} for zero-shot audio quality assessment by finding high- and low-quality anchor points in the CLAP embedding space.

Despite the benefits of using SSL model features in speech quality assessment, they significantly increase model size and surface features unrelated to speech quality. To reduce model size, we investigate  distillation and pruning methods.

In the remainder of this paper, \cref{sec:task} formalizes the problem of non-intrusive computational speech quality assessment. \Cref{sec:data} provides an overview of the datasets used in our study. \Cref{sec:impr}, \cref{sec:distill}, and \cref{sec:taylorprune} describe improvements for training an XLS-R-based model on these data, the method of distilling the teacher model into more compact student models using unlabeled data, and the method of data-driven pruning, respectively. Baselines are described in \Cref{sec:baselines}, and experimental results are discussed in \Cref{sec:results}. Finally, \Cref{sec:concl} summarizes the results and insights. We make the weights of a selected distilled model available$^\textrm{1}$.

\section{Non-Intrusive Speech Quality Assessment}\label{sec:task}
In subjective evaluations, such as those based on ITU-T standards P.800, P.808, P.835, and P.804, multiple raters assess the overall quality (and other aspects in the case of P.835 and P.804) of a degraded speech signal $x$, scoring it on a scale from 1 (bad) to 5 (excellent). The mean opinion score $\mathrm{MOS}$ is the average of these scores across raters. A non-intrusive speech quality assessment model $f: x \rightarrow \widehat{\mathrm{MOS}}$ predicts this average score directly from the speech signal, estimating perceived quality without a reference signal. The per-dataset Pearson's correlation coefficient (PCC) is commonly used for evaluation.

\section{Labeled MOS Datasets}\label{sec:data}
For the experiments described in this paper, we use a large-scale compilation of MOS-labeled speech. Across all datasets, MOS labels were collected from different numbers of raters using various procedures, but all labels reflect \textit{overall} quality. Numbers in parentheses for each dataset denote number of clips in the (training $|$ validation $|$ test) splits.

\subsection{Data Used for Training, Validation, and Testing}
\begin{itemize}[leftmargin=4mm]
\item \textbf{PSTN} \cite{mittag2020} (\numprint{57709} $|$ 1000 (custom split) $|$ 1040 clips) contains English speech signals with degradations from live calls through public switched telphone networks. Uses the training data provided with the first Deep Noise Suppression challenge as input speech/noise data. 

\item \textbf{Tencent} \cite{yi2022}  (\numprint{10563} $|$ 1000 (custom split) $|$ 2898 clips) is part of the ConferencingSpeech 2022 challenge data. Contains simulated and live VoIP-typical degradations applied to speech from three public Chinese datasets (Magic data, ST Mandarin, and AIshell-100h). 

\item \textbf{NISQA (simulated)} \cite{mittag2021} (\numprint{10000} $|$ 2500 $|$ -- clips) contains simulated degradations on English speech from Librivox, TSP speech database, `Crowdsourced high-quality UK and Ireland English Dialect speech data set'', and AusTalk. 

\item  \textbf{NISQA (live)} \cite{mittag2021}  (1020 $|$ 200 $|$ -- clips) contains recordings of live VoIP calls using the Librivox speech data from the first DNS challenge (English). 

\item \textbf{VoiceMOS challenge 2022 English} \cite{huang2022} (4973 $|$ 1065 $|$ 1065 clips) contains synthetic speech from multiple Blizzard and Voice Conversion challenges. 

\item \textbf{VoiceMOS challenge 2022 Chinese} \cite{huang2022}  (674 $|$ 136 $|$ 539 clips) contains synthetic Chinese speech from the 2019 Blizzard challenge. We merged the ``Unlabeled'' dataset split, whose labels had later been added by the challenge organizers, into the training split. 
\end{itemize}

Additonally, we used internal MOS-labeled data from submissions to different challenges organized by Microsoft. For the internal MOS-labeled challenge submission data, we created validation splits by randomly setting aside 1/4 of the submission systems and 1/4 of the input signals.
\begin{itemize}[leftmargin=4mm]
\item \textbf{Deep Noise Suppression (DNS) challenge 2 submissions} \cite{reddy2021} (8918 $|$ 1041 $|$ -- clips) contains noisy speech in diverse languages enhanced by the submitted methods. 
\item \textbf{Packet Loss Concealment (PLC) challenge 2022 submissions} \cite{diener2022} (9412 $|$ 958 $|$ -- clips) contains speech with simulated packet loss concealed by the submitted methods.
\item \textbf{Signal Improvement (SIG) challenge 2023 submissions} \cite{cutler2023} (5250 $|$ 625 $|$ -- clips) contains speech with various degradations enhanced by the submitted methods.
\end{itemize}

\subsection{Additional Data Used for Testing}
\begin{itemize}[leftmargin=4mm]
\item \textbf{TUB} \cite{yi2022} (434 clips) has been released as part of the ConferncingSpeech 2022 test data. It contains conversational speech with synthetic degradations.
\item \textbf{NISQA Forensic} \cite{mittag2021} (240 clips) contains speech from the Forensic Voice Comparison Databases \cite{morrison2012} with degradations including simulated noise, packet loss, and clipping and real (Whatsapp, Zoom, and Discord).
\item \textbf{NISQA Livetalk} \cite{mittag2021} (232 clips) contains German conversational speech in real acoustic and background noise conditions with real VoIP transmission (Skype/Facebook). 
\item \textbf{NISQA P.501} \cite{mittag2021} (240 clips) contains simulated degradations and real degradations from transmission unsing VoIP services (Skype, Zoom, WhatsApp) and mobile network on English speech taken from the ITU-T P.501 dataset.
\item \textbf{TCD-VoIP} \cite{harte2015} (240 clips) contains simulated degradations (noise, competing speaker, clipping, chopping/dropouts, echo) on English speech from the TSP speech database.
\item \textbf{Genspeech} \cite{jassim2020} (160 clips) contains speech from the NTT Multilingual Speech Database coded and reconstructed using different neural and traditional codecs. The original MUSHRA scores on a scale from 0 to 100 have been rescaled to match the MOS scale (1 to 5).
\item \textbf{Blizzard challenge 2023 submissions} \cite{perrotin2023} contains synthetic French speech from a text-to-speech (TTS) task (840 clips) and a speaker adaptation (SA) task (578 clips).
\end{itemize}

The MOS-rated \textbf{DNS challenge 3} \cite{reddy2021_2} (\numprint{16800} clips) (and the MOS-rated fullband track submissions, denoted by DNS\,3~FB (4200 clips)), \textbf{PLC challenge 2024} \cite{diener2024} (\numprint{11010} clips), and \textbf{SIG challenge 2024} \cite{ristea2024} (5100 clips) submissions, i.e., the respective challenges following up on the challenge MOS datasets used for training and validation, were also used for testing. We removed the fraction of input signals overlapping with the SIG challenge 2023 dataset from the SIG challenge 2024 test dataset.

In total, \numprint{108229} MOS-labeled speech clips were used for training, 8534 for validation, and 46050 for testing.

\section{Proposed Method}\label{sec:method}

\subsection{Improvements for XLS-R-Based Speech Quality Assessment}\label{sec:impr}
Based on a comparison of state-of-the-art speech quality assessment models, we use XLS-R-SQA \cite{tamm2023}, referred to as \textit{Orig. XLS-R(2B)-L10+T}, as our base model due to its superior generalization. Following \cite{tamm2023}, we used activations from the 10th layer of the 2-billion-parameter \mbox{XLS-R} model as input features for the MOS prediction head. The MOS prediction head consists of a four-layer transformer with a hidden dimension of 32 and 4 attention heads, followed by attention pooling. The model was retrained on the data described in \cref{sec:data} for \numprint{72000} update steps with a batch size of 12. A sampling strategy was applied, limiting the probability of sampling from datasets larger than 7000 clips. The XLS-R embedding model was not trained/fine-tuned. As loss function, we used the mean squared error between predicted MOS and ground truth MOS. The AdamW optimizer \cite{loshchilov2018} with learning rate 1e-4 was used. Every 3000 update steps, the model was validated on the validation datasets. The model yielding the lowest weighted mean squared error during validation was selected. The retrained model is denoted by \textit{XLS-R(2B)-L10+T}. To account for potential dataset bias within the diverse training datasets, we introduced per-dataset learnable scale and shift parameters applied to the logits prior to the sigmoid output activation, similar to \cite{mittag2021_2}. After training, optimal ``universal'' scale and shift parameters were determined through grid search. This model is denoted as \textit{XLS-R(2B)-L10+T(BA)}.

\subsection{Distillation into Compact Models}\label{sec:distill}

\begin{figure}
\centering
\hspace*{3.5mm}
\includegraphics[width=1.0\linewidth]{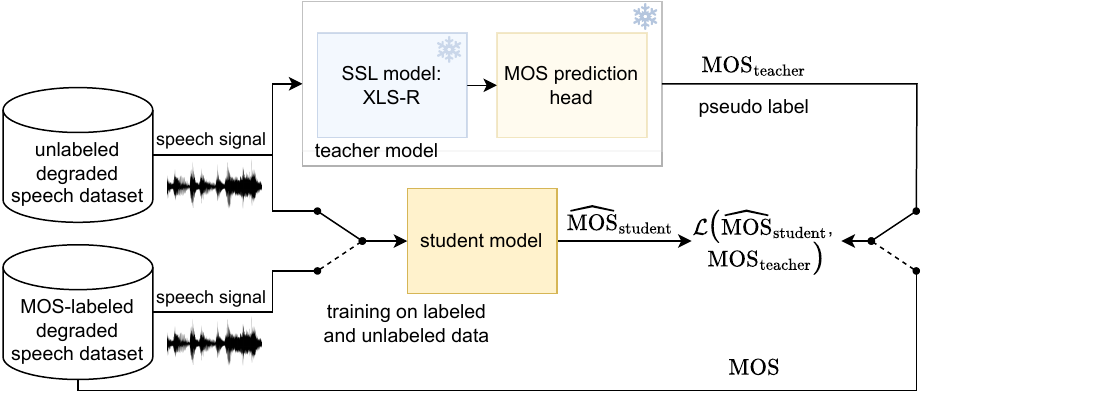}
\vspace*{-3.5mm}
\caption{XLS-R-based speech quality assessment and its usage as a teacher model for distillation using unlabeled speech.}\label{fig:distillfull}
\vspace*{-2mm}
\end{figure}

The XLS-R embeddings used for speech quality assessment are derived from a general pretext task, also capturing elements of speech unrelated to speech quality. To transfer the benefits of XLS-R to more compact models, we propose a straightforward distillation approach. A student model is trained to replicate the output of the XLS-R-based teacher model \textit{XLS-R(2B)-L10+T(BA)} using a large dataset of unlabeled speech, cf. \cref{fig:distillfull}.

\begin{figure}
\centering
\vspace*{-2mm}
\includegraphics[width=0.96\linewidth]{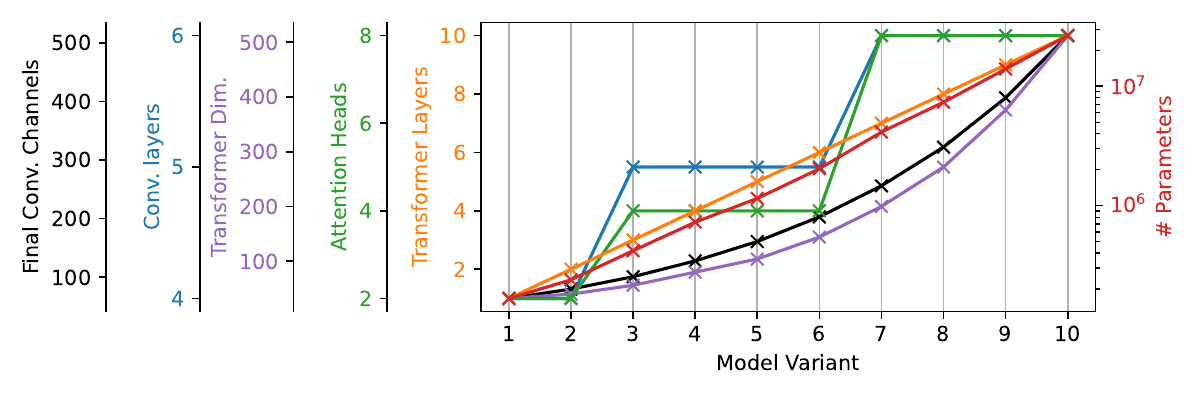}
\vspace*{-3mm}
\caption{Resulting model sizes for student model variants.} \label{fig:hyperparams}
\vspace*{-6mm}
\end{figure}

The unlabeled data used for distillation include generated speech and typical VoIP degradations. Specifically, 15\% of the unlabeled training samples were generated using 17 different public text-to-speech models \cite{coqui} in English, German, French, Spanish, and Portuguese, with text sourced from the MultiSubs dataset \cite{wang2022}. The remaining 85\% consisted of real speech signals subjected to degradations. These signals were based on clean speech ($\sim$2600\,h), noise ($\sim$180\,h), and room impulse responses from the 4th DNS challenge training dataset \cite{dubey2022}. In 33\% of the real-speech scenarios, a random room impulse response was applied. In 67\%, noise was added with a Gaussian-distributed SNR (mean 10 dB, variance 10 dB). Next, one of the following was uniformly selected: DeepFilterNet3 \cite{schroter2023}, NSNet2 \cite{braun2020}, Demucs \cite{defossez2020}, Spectral Gating \cite{tim_sainburg_2019_3243139}, or no denoising. To simulate generative enhancement, HiFiGAN \cite{kong2020} or LPCNet \cite{valin2019} vocoders were applied in 33\% of cases. Finally, in 50\% of the cases, coding was applied, randomly selected from Opus \cite{valin2013high}, EnCodec \cite{defossez2023high}, or a 4\,kHz bandwidth limitation.

For the student model, which does not use XLS-R features, we employed a convolutional transformer architecture. Like XLS-R, our models use signal components up to 8\,kHz, so all speech signals were resampled to 16\,kHz. The input consists of the real and imaginary parts of complex compressed spectrogram features \cite{braun2022} with a 320-sample window length, 160-sample hop length, and a Hann window. These features are processed by a convolutional neural network with initial layers having kernel size (3,3) and stride (1,1) along the (frequency, time) axes. Subsequent layers use kernel size (3,3) with stride (2,1), and the final layer uses stride (2,2) to downsample along the time axis before transformer encoding. The first two convolutional layers have 64 channels, doubling in later layers until reaching a defined maximum number of channels. All convolutions were followed by a LeakyReLU activation function with the slope set at 0.1 for negative input, We varied the dimension parameters to create 10 model variants with different model sizes as shown in \cref{fig:hyperparams}.

For distillation, we mixed in labeled clips and corresponding ground truth labels into the unlabeled data at a percentage $p$. A grid search was conducted over $p$ values of 0\%, 10\%, 20\%, 40\%, and 80\%, revealing that 20\% was the optimal mix-in percentage. The student model was trained on pseudo-labels or ground truth labels with a batch size of 20. Ground truth labels were scaled and shifted in the logits domain using the inverse transform from the teacher model’s universal to per-dataset scale and shift parameters. Mean squared error was used as the loss function, with the AdamW optimizer (learning rate: 2e-5). The models were trained for \numprint{250000} steps and validated every 5000, selecting the best model based on validation mean squared error. We denote these models by \textit{XLS-R-DistillMOS}.

\subsection{Importance-Based Pruning} \label{sec:taylorprune}
Besides distillation, we investigate pruning as a strategy to reduce model size. We employed the data-driven pruning method proposed in \cite{molchanov2019}. For each parameter $w$, an importance score $\mathcal{I}_w$ is approximated by its first-order Taylor expansion:
\vspace*{-1.25mm}
\begin{equation}
\mathcal{I}_w = \left(\mathcal{L} - \mathcal{L}_{w=0}\right)^2 \approx \left(\frac{d \mathcal{L}}{d w} w\right)^2,
\vspace*{-1.25mm}
\end{equation}
where $\mathcal{L}$ denotes the loss function evaluated with the full model and $\mathcal{L}_{w=0}$ denotes the loss function with parameter $w$ set to 0.

In each pruning step, 0.5\% of weights of the XLS-R embedding model with the lowest importance scores were removed. The MOS prediction head was not pruned. After each pruning step, the model was fine-tuned for 30 update steps on the labeled data described in \cref{sec:data}. Fine-tuning was performed parallel to importance score aggregation. The AdamW optimizer with learning rate 2e-5 and batch size 20 was used for fine-tuning. Importance scores were recursively averaged using a smoothing factor of 0.9. Pruned model checkpoints were saved at multiple points during the pruning process to enable evaluation at different model sizes. We denote these models by \textit{XLS-R-PruneMOS}.


We also explored pruning and fine-tuning based on unlabeled data and pseudo-labels from the unpruned model. However, this approach did not result in improved pruning performance.

\section{Compared Baselines}\label{sec:baselines}

\newcommand{\mysaturation}{0.5}
\newcommand{\mybrightness}{0.9}
\newcommand{\myheaderheight}{1.4cm}

\renewcommand*{\arraystretch}{0.8}
\setlength\tabcolsep{4.35pt}
\begin{table*}[ht!]
\scriptsize
\centering
\caption{Pearson correlation coefficient on test datasets for baselines, teacher model, and selected distilled and pruned models.}\label{tab:restable}
\vspace{-2mm}
\begin{tabular}{p{3.0cm} c | c c c c c c c c c c c c c c c c | c c}
\toprule
 & \multicolumn{9}{c}{VoIP} & \multicolumn{4}{c}{TTS / Voice Convers.}  & & &  \\
 \cmidrule(lr){3-10} 
\cmidrule(lr){11-14} 
 & \rotatebox[]{90}{\parbox[c]{\myheaderheight}{\mbox{\# params (M)}}} 
 & \rotatebox[]{90}{\parbox[c]{\myheaderheight}{PSTN}} & \rotatebox[]{90}{\parbox[c]{\myheaderheight}{Tencent}} & \rotatebox[]{90}{\parbox[c]{\myheaderheight}{TUB}} & \rotatebox[]{90}{\parbox[c]{\myheaderheight}{NISQA For}} & \rotatebox[]{90}{\parbox[c]{\myheaderheight}{NISQA Livet.}} & \rotatebox[]{90}{\parbox[c]{\myheaderheight}{NISQA P.501}} & \rotatebox[]{90}{\parbox[c]{\myheaderheight}{TCD-VoIP}} & \rotatebox[]{90}{\parbox[c]{\myheaderheight}{Genspeech}} & \rotatebox[]{90}{\parbox[c]{\myheaderheight}{VoiceMOS EN}} & \rotatebox[]{90}{\parbox[c]{\myheaderheight}{VoiceMOS CN}} & \rotatebox[]{90}{\parbox[c]{\myheaderheight}{Blizzard 2023 (FR) TTS}} & \rotatebox[]{90}{\parbox[c]{\myheaderheight}{Blizzard 2023 (FR) SA}} & \rotatebox[]{90}{\parbox[c]{\myheaderheight}{DNS 3}}  & \rotatebox[]{90}{\parbox[c]{\myheaderheight}{DNS 3 FB}}  & \rotatebox[]{90}{\parbox[c]{\myheaderheight}{PLC 2024}} & \rotatebox[]{90}{\parbox[c]{\myheaderheight}{Sig 2024}}  & \rotatebox[]{90}{\parbox[c]{\myheaderheight}{weighted mean}}  & \rotatebox[]{90}{\parbox[c]{\myheaderheight}{unweighted mean}}\\
\midrule
\#~clips & & \scriptsize 1040  & \scriptsize 2898 & \scriptsize 434  & \scriptsize 240 & \scriptsize 232& \scriptsize 240 & \scriptsize 384 & \scriptsize 160 & \scriptsize 1065& \scriptsize 539& \scriptsize 840&  \scriptsize 578 & \scriptsize 16.8k& \scriptsize 4500& \scriptsize 11.0k & \scriptsize 5100 & \\ 
\midrule
DNSMOS P.808 \cite{reddy2021_3}& 0.05 &\cellcolor[hsb]{0.0880328, \mysaturation, \mybrightness}0.49 & \cellcolor[hsb]{0.243441, \mysaturation, \mybrightness}0.82 & \cellcolor[hsb]{0.129135, \mysaturation, \mybrightness}0.57 & \cellcolor[hsb]{0.11039, \mysaturation, \mybrightness}0.53 & \cellcolor[hsb]{0.195102, \mysaturation, \mybrightness}0.71 & \cellcolor[hsb]{0.175684, \mysaturation, \mybrightness}0.67 & \cellcolor[hsb]{0.198968, \mysaturation, \mybrightness}0.72 & \cellcolor[hsb]{0.194157, \mysaturation, \mybrightness}0.71 & \cellcolor[hsb]{0.146919, \mysaturation, \mybrightness}0.61 & \cellcolor[hsb]{0.0747182, \mysaturation, \mybrightness}0.46 & \cellcolor[hsb]{0.0399326, \mysaturation, \mybrightness}0.38 & \cellcolor[hsb]{0.0170863, \mysaturation, \mybrightness}0.34 & \cellcolor[hsb]{0.195254, \mysaturation, \mybrightness}0.71 & \cellcolor[hsb]{0.202872, \mysaturation, \mybrightness}0.73 & \cellcolor[hsb]{0.166332, \mysaturation, \mybrightness}0.65 & \cellcolor[hsb]{0.0760164, \mysaturation, \mybrightness}0.46 & \cellcolor[hsb]{0.167526, \mysaturation, \mybrightness}0.66 & \cellcolor[hsb]{0.140878, \mysaturation, \mybrightness}0.60\\
DNSMOS P.835 \cite{reddy2022}& 0.17 &\cellcolor[hsb]{0.102001, \mysaturation, \mybrightness}0.52 & \cellcolor[hsb]{0.219009, \mysaturation, \mybrightness}0.76 & \cellcolor[hsb]{0.0801156, \mysaturation, \mybrightness}0.47 & \cellcolor[hsb]{0.113454, \mysaturation, \mybrightness}0.54 & \cellcolor[hsb]{0.142392, \mysaturation, \mybrightness}0.60 & \cellcolor[hsb]{0.165451, \mysaturation, \mybrightness}0.65 & \cellcolor[hsb]{0.118031, \mysaturation, \mybrightness}0.55 & \cellcolor[hsb]{0.207648, \mysaturation, \mybrightness}0.74 & \cellcolor[hsb]{0.0872386, \mysaturation, \mybrightness}0.49 & \cellcolor[hsb]{0.0837527, \mysaturation, \mybrightness}0.48 & \cellcolor[hsb]{0, \mysaturation, \mybrightness}0.12 & \cellcolor[hsb]{0, \mysaturation, \mybrightness}0.25 & \cellcolor[hsb]{0.25678, \mysaturation, \mybrightness}\textbf{0.84} & \cellcolor[hsb]{0.256372, \mysaturation, \mybrightness}\textbf{0.84} & \cellcolor[hsb]{0.172334, \mysaturation, \mybrightness}0.67 & \cellcolor[hsb]{0.0692254, \mysaturation, \mybrightness}0.45 & \cellcolor[hsb]{0.189043, \mysaturation, \mybrightness}0.70 & \cellcolor[hsb]{0.123032, \mysaturation, \mybrightness}0.56\\
NISQA \cite{mittag2021}& 0.25 &\cellcolor[hsb]{0.0985897, \mysaturation, \mybrightness}0.51 & \cellcolor[hsb]{0.218576, \mysaturation, \mybrightness}0.76 & \cellcolor[hsb]{0.218666, \mysaturation, \mybrightness}0.76 & \cellcolor[hsb]{0.278445, \mysaturation, \mybrightness}0.89 & \cellcolor[hsb]{0.225483, \mysaturation, \mybrightness}0.78 & \cellcolor[hsb]{0.284601, \mysaturation, \mybrightness}0.90 & \cellcolor[hsb]{0.25554, \mysaturation, \mybrightness}0.84 & \cellcolor[hsb]{0.267773, \mysaturation, \mybrightness}0.87 & \cellcolor[hsb]{0.151022, \mysaturation, \mybrightness}0.62 & \cellcolor[hsb]{0.0629489, \mysaturation, \mybrightness}0.43 & \cellcolor[hsb]{0.087378, \mysaturation, \mybrightness}0.49 & \cellcolor[hsb]{0, \mysaturation, \mybrightness}0.30 & \cellcolor[hsb]{0.152588, \mysaturation, \mybrightness}0.62 & \cellcolor[hsb]{0.151053, \mysaturation, \mybrightness}0.62 & \cellcolor[hsb]{0.175524, \mysaturation, \mybrightness}0.67 & \cellcolor[hsb]{0.101601, \mysaturation, \mybrightness}0.52 & \cellcolor[hsb]{0.154621, \mysaturation, \mybrightness}0.63 & \cellcolor[hsb]{0.170551, \mysaturation, \mybrightness}0.66\\
Ta-SQUIM PESQ \cite{kumar2023}& 7.4 &\cellcolor[hsb]{0.0668295, \mysaturation, \mybrightness}0.44 & \cellcolor[hsb]{0.211199, \mysaturation, \mybrightness}0.75 & \cellcolor[hsb]{0.182164, \mysaturation, \mybrightness}0.69 & \cellcolor[hsb]{0.199117, \mysaturation, \mybrightness}0.72 & \cellcolor[hsb]{0.140516, \mysaturation, \mybrightness}0.60 & \cellcolor[hsb]{0.25521, \mysaturation, \mybrightness}0.84 & \cellcolor[hsb]{0.150403, \mysaturation, \mybrightness}0.62 & \cellcolor[hsb]{0.237686, \mysaturation, \mybrightness}0.80 & \cellcolor[hsb]{0.192478, \mysaturation, \mybrightness}0.71 & \cellcolor[hsb]{0.0775695, \mysaturation, \mybrightness}0.46 & \cellcolor[hsb]{0, \mysaturation, \mybrightness}0.16 & \cellcolor[hsb]{0.00399591, \mysaturation, \mybrightness}0.31 & \cellcolor[hsb]{0.166178, \mysaturation, \mybrightness}0.65 & \cellcolor[gray]{0.8} N/A & \cellcolor[hsb]{0.199509, \mysaturation, \mybrightness}0.72 & \cellcolor[hsb]{0.112283, \mysaturation, \mybrightness}0.54 & \cellcolor[hsb]{0.162512, \mysaturation, \mybrightness}0.64 & \cellcolor[hsb]{0.142043, \mysaturation, \mybrightness}0.60\\
UTMOS \cite{saeki2022} & 103 &\cellcolor[hsb]{0.145653, \mysaturation, \mybrightness}0.61 & \cellcolor[hsb]{0.183873, \mysaturation, \mybrightness}0.69 & \cellcolor[hsb]{0.193294, \mysaturation, \mybrightness}0.71 & \cellcolor[hsb]{0.236364, \mysaturation, \mybrightness}0.80 & \cellcolor[hsb]{0.221838, \mysaturation, \mybrightness}0.77 & \cellcolor[hsb]{0.261316, \mysaturation, \mybrightness}0.85 & \cellcolor[hsb]{0.240883, \mysaturation, \mybrightness}0.81 & \cellcolor[hsb]{0.292542, \mysaturation, \mybrightness}0.92 & \cellcolor[hsb]{0.274638, \mysaturation, \mybrightness}\textbf{0.88} & \cellcolor[hsb]{0.147182, \mysaturation, \mybrightness}0.61 & \cellcolor[hsb]{0.0910792, \mysaturation, \mybrightness}0.49 & \cellcolor[hsb]{0, \mysaturation, \mybrightness}0.20 & \cellcolor[hsb]{0.139044, \mysaturation, \mybrightness}0.59 & \cellcolor[hsb]{0.145128, \mysaturation, \mybrightness}0.61 & \cellcolor[hsb]{0.23728, \mysaturation, \mybrightness}0.80 & \cellcolor[hsb]{0.150275, \mysaturation, \mybrightness}0.62 & \cellcolor[hsb]{0.171005, \mysaturation, \mybrightness}0.66 & \cellcolor[hsb]{0.182224, \mysaturation, \mybrightness}0.69\\
Orig. XLS-R(2B)-L10+T \cite{tamm2023}& 478 &\cellcolor[hsb]{0.12204, \mysaturation, \mybrightness}0.56 & \cellcolor[hsb]{0.30722, \mysaturation, \mybrightness}0.95 & \cellcolor[hsb]{0.254512, \mysaturation, \mybrightness}0.84 & \cellcolor[hsb]{0.293148, \mysaturation, \mybrightness}\textbf{0.92} & \cellcolor[hsb]{0.283415, \mysaturation, \mybrightness}0.90 & \cellcolor[hsb]{0.305379, \mysaturation, \mybrightness}\textbf{0.95} & \cellcolor[hsb]{0.256744, \mysaturation, \mybrightness}0.84 & \cellcolor[hsb]{0.308886, \mysaturation, \mybrightness}\textbf{0.96} & \cellcolor[hsb]{0.21293, \mysaturation, \mybrightness}0.75 & \cellcolor[hsb]{0.181979, \mysaturation, \mybrightness}0.69 & \cellcolor[hsb]{0.146838, \mysaturation, \mybrightness}\textbf{0.61} & \cellcolor[hsb]{0.134708, \mysaturation, \mybrightness}0.59 & \cellcolor[hsb]{0.226781, \mysaturation, \mybrightness}0.78 & \cellcolor[hsb]{0.226991, \mysaturation, \mybrightness}0.78 & \cellcolor[hsb]{0.24211, \mysaturation, \mybrightness}0.81 & \cellcolor[hsb]{0.182231, \mysaturation, \mybrightness}0.69 & \cellcolor[hsb]{0.226605, \mysaturation, \mybrightness}0.78 & \cellcolor[hsb]{0.23037, \mysaturation, \mybrightness}0.79\\  
\midrule
XLS-R(2B)-L10+T& 478 &\cellcolor[hsb]{0.144882, \mysaturation, \mybrightness}0.61 & \cellcolor[hsb]{0.31248, \mysaturation, \mybrightness}0.96 & \cellcolor[hsb]{0.260545, \mysaturation, \mybrightness}0.85 & \cellcolor[hsb]{0.285747, \mysaturation, \mybrightness}0.91 & \cellcolor[hsb]{0.291007, \mysaturation, \mybrightness}\textbf{0.92} & \cellcolor[hsb]{0.300993, \mysaturation, \mybrightness}0.94 & \cellcolor[hsb]{0.285068, \mysaturation, \mybrightness}\textbf{0.90} & \cellcolor[hsb]{0.298258, \mysaturation, \mybrightness}0.93 & \cellcolor[hsb]{0.271166, \mysaturation, \mybrightness}0.88 & \cellcolor[hsb]{0.283784, \mysaturation, \mybrightness}0.90 & \cellcolor[hsb]{0.125459, \mysaturation, \mybrightness}0.57 & \cellcolor[hsb]{0.17227, \mysaturation, \mybrightness}\textbf{0.67} & \cellcolor[hsb]{0.231956, \mysaturation, \mybrightness}0.79 & \cellcolor[hsb]{0.232177, \mysaturation, \mybrightness}0.79 & \cellcolor[hsb]{0.251984, \mysaturation, \mybrightness}0.83 & \cellcolor[hsb]{0.180646, \mysaturation, \mybrightness}0.68 & \cellcolor[hsb]{0.234904, \mysaturation, \mybrightness}0.80 & \cellcolor[hsb]{0.245526, \mysaturation, \mybrightness}0.82\\
XLS-R(2B)-L10+T(BA)& 478 &\cellcolor[hsb]{0.1574, \mysaturation, \mybrightness}0.63 & \cellcolor[hsb]{0.313912, \mysaturation, \mybrightness}\textbf{0.97} & \cellcolor[hsb]{0.263927, \mysaturation, \mybrightness}\textbf{0.86} & \cellcolor[hsb]{0.29229, \mysaturation, \mybrightness}0.92 & \cellcolor[hsb]{0.289518, \mysaturation, \mybrightness}0.91 & \cellcolor[hsb]{0.297374, \mysaturation, \mybrightness}0.93 & \cellcolor[hsb]{0.27912, \mysaturation, \mybrightness}0.89 & \cellcolor[hsb]{0.293647, \mysaturation, \mybrightness}0.92 & \cellcolor[hsb]{0.27218, \mysaturation, \mybrightness}0.88 & \cellcolor[hsb]{0.28491, \mysaturation, \mybrightness}\textbf{0.90} & \cellcolor[hsb]{0.133706, \mysaturation, \mybrightness}0.58 & \cellcolor[hsb]{0.142254, \mysaturation, \mybrightness}0.60 & \cellcolor[hsb]{0.24466, \mysaturation, \mybrightness}0.82 & \cellcolor[hsb]{0.24837, \mysaturation, \mybrightness}0.83 & \cellcolor[hsb]{0.254191, \mysaturation, \mybrightness}\textbf{0.84} & \cellcolor[hsb]{0.188468, \mysaturation, \mybrightness}\textbf{0.70} & \cellcolor[hsb]{0.242615, \mysaturation, \mybrightness}\textbf{0.81} & \cellcolor[hsb]{0.247245, \mysaturation, \mybrightness}\textbf{0.82}\\
\midrule
labeled-only baseline v7& 4.3 &\cellcolor[hsb]{0.111471, \mysaturation, \mybrightness}0.54 & \cellcolor[hsb]{0.29797, \mysaturation, \mybrightness}0.93 & \cellcolor[hsb]{0.210175, \mysaturation, \mybrightness}0.75 & \cellcolor[hsb]{0.262123, \mysaturation, \mybrightness}0.86 & \cellcolor[hsb]{0.232649, \mysaturation, \mybrightness}0.79 & \cellcolor[hsb]{0.274892, \mysaturation, \mybrightness}0.88 & \cellcolor[hsb]{0.194517, \mysaturation, \mybrightness}0.71 & \cellcolor[hsb]{0.228627, \mysaturation, \mybrightness}0.78 & \cellcolor[hsb]{0.199013, \mysaturation, \mybrightness}0.72 & \cellcolor[hsb]{0.268273, \mysaturation, \mybrightness}0.87 & \cellcolor[hsb]{0, \mysaturation, \mybrightness}0.24 & \cellcolor[hsb]{0, \mysaturation, \mybrightness}0.26 & \cellcolor[hsb]{0.20943, \mysaturation, \mybrightness}0.74 & \cellcolor[hsb]{0.215014, \mysaturation, \mybrightness}0.76 & \cellcolor[hsb]{0.226504, \mysaturation, \mybrightness}0.78 & \cellcolor[hsb]{0.0974101, \mysaturation, \mybrightness}0.51 & \cellcolor[hsb]{0.19881, \mysaturation, \mybrightness}0.72 & \cellcolor[hsb]{0.186229, \mysaturation, \mybrightness}0.70\\
\textbf{XLS-R-DistillMOS v7} & 4.3 &\cellcolor[hsb]{0.129802, \mysaturation, \mybrightness}0.58 & \cellcolor[hsb]{0.305358, \mysaturation, \mybrightness}0.95 & \cellcolor[hsb]{0.21842, \mysaturation, \mybrightness}0.76 & \cellcolor[hsb]{0.241278, \mysaturation, \mybrightness}0.81 & \cellcolor[hsb]{0.263252, \mysaturation, \mybrightness}0.86 & \cellcolor[hsb]{0.282645, \mysaturation, \mybrightness}0.90 & \cellcolor[hsb]{0.255554, \mysaturation, \mybrightness}0.84 & \cellcolor[hsb]{0.244588, \mysaturation, \mybrightness}0.82 & \cellcolor[hsb]{0.242879, \mysaturation, \mybrightness}0.82 & \cellcolor[hsb]{0.253945, \mysaturation, \mybrightness}0.84 & \cellcolor[hsb]{0, \mysaturation, \mybrightness}0.22 & \cellcolor[hsb]{0.011316, \mysaturation, \mybrightness}0.32 & \cellcolor[hsb]{0.229078, \mysaturation, \mybrightness}0.79 & \cellcolor[hsb]{0.236845, \mysaturation, \mybrightness}0.80 & \cellcolor[hsb]{0.240868, \mysaturation, \mybrightness}0.81 & \cellcolor[hsb]{0.139298, \mysaturation, \mybrightness}0.60 & \cellcolor[hsb]{0.218859, \mysaturation, \mybrightness}0.76 & \cellcolor[hsb]{0.2036, \mysaturation, \mybrightness}0.73\\
\textbf{XLS-R-PruneMOS 29\%} & 139$^\dagger$ &\cellcolor[hsb]{0.170719, \mysaturation, \mybrightness}\textbf{0.66} & \cellcolor[hsb]{0.310566, \mysaturation, \mybrightness}0.96 & \cellcolor[hsb]{0.258371, \mysaturation, \mybrightness}0.85 & \cellcolor[hsb]{0.284047, \mysaturation, \mybrightness}0.90 & \cellcolor[hsb]{0.274275, \mysaturation, \mybrightness}0.88 & \cellcolor[hsb]{0.281749, \mysaturation, \mybrightness}0.90 & \cellcolor[hsb]{0.256116, \mysaturation, \mybrightness}0.84 & \cellcolor[hsb]{0.298242, \mysaturation, \mybrightness}0.93 & \cellcolor[hsb]{0.265593, \mysaturation, \mybrightness}0.86 & \cellcolor[hsb]{0.282914, \mysaturation, \mybrightness}0.90 & \cellcolor[hsb]{0.10425, \mysaturation, \mybrightness}0.52 & \cellcolor[hsb]{0.0541543, \mysaturation, \mybrightness}0.41 & \cellcolor[hsb]{0.236546, \mysaturation, \mybrightness}0.80 & \cellcolor[hsb]{0.241611, \mysaturation, \mybrightness}0.81 & \cellcolor[hsb]{0.253016, \mysaturation, \mybrightness}0.84 & \cellcolor[hsb]{0.177234, \mysaturation, \mybrightness}0.68 & \cellcolor[hsb]{0.235309, \mysaturation, \mybrightness}0.80 & \cellcolor[hsb]{0.234338, \mysaturation, \mybrightness}0.80\\
\bottomrule
\end{tabular}
\vspace*{-5mm}
\end{table*}

\subsection{Labeled-Data-Only Baseline Models}
We trained baseline models using only labeled data, employing the same convolutional transformer architecture and training procedure as the student models with distillation.

\subsection{Na\"\i ve Magnitude-Based Pruning}
As a pruning baseline, we applied a simple magnitude-based pruning \cite{gale2019} to the XLS-R model. As with importance-based pruning, in each iteration, we removed the 0.5\% of weights of the XLS-R embedding model with the lowest magnitude from each weight matrix, and the MOS prediction head was not pruned. For convolutional layers, the L1 norm of kernels was considered. Pruning was done separately for each weight/kernel matrix. After each pruning step, the model was fine-tuned on labeled MOS data for 30 update steps using the AdamW optimizer with learning rate 2e-5 and batch size 20. Checkpoints were saved at multiple points during the pruning process.

\subsection{Other State-Of-The-Art Speech Quality Assessment Models}
We compare the proposed models to DNSMOS P.808 \cite{reddy2021_3}, DNSMOS P.835 \cite{reddy2022}, NISQA \cite{mittag2021}, TorchaudioSQUIM-PESQ \cite{kumar2023} (which had been trained on unlabeled data using PESQ as a teacher), and to the original XLS-R SQA model \cite{tamm2023}. To also compare to a model for quality assessment of generated speech, we included UTMOS \cite{saeki2022}.

\section{Experimental Results and Discussion}\label{sec:results}
\begin{figure}
\centering
\vspace*{-1mm}
\includegraphics[width=0.93\linewidth]{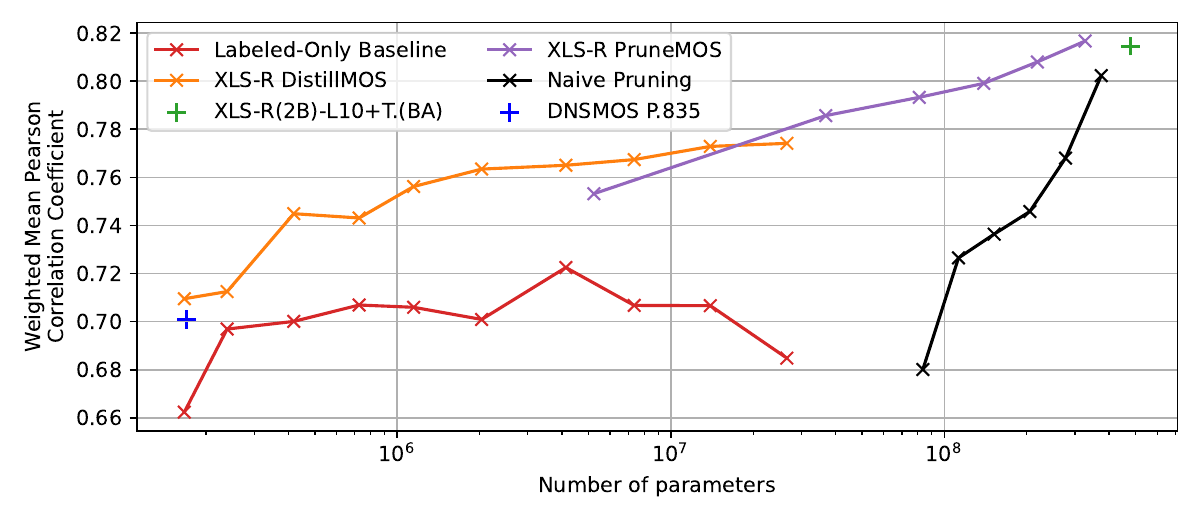}
\vspace*{-3mm}
\caption{Weighted average Pearson correlation coefficient as a function of model size with different approaches$^\dagger$.}\label{fig:resnumparams}
\vspace*{-5mm}
\end{figure}

\Cref{tab:restable} provides an overview of model performance based on the Pearson correlation coefficient with ground truth MOS labels as well as the weighted (by number of clips) and unweighted mean of the correlation coefficient across test datasets. The first section of the table evaluates state-of-the-art speech quality assessment models. DNSMOS performs well on the DNS 3 dataset but generalizes poorly to others. NISQA excels on the NISQA datasets and generalizes more broadly. TorchaudioSQUIM PESQ only achieves moderate correlation on most datasets. UTMOS performs strongly on the VoiceMOS EN dataset and even generalizes to some VoIP data. \textit{Orig. XLS-R(2B)-L10+T} achieves the highest overall correlation and generalization. Although none of the models generalize well to out-of-distribution synthetic speech data, \textit{Orig. XLS-R(2B)-L10+T} achieves highest correlation on these.

\blfootnote{$^\dagger$
Sparse indices count as half parameters (representable by 16-bit integers).}

The second section highlights the retrained XLS-R models. Retraining on diverse data and using learned output transforms improves correlation scores. Notably, the correlation on out-of-domain synthetic speech data, i.e., the Blizzard datasets, did not improve by including synthetic data from other domains. This indicates that the relevant features of speech quality vary between different synthetic speech datasets.

\Cref{fig:resnumparams} shows the average (weighted by number of clips) Pearson correlation for labeled-only baseline, distilled, and pruned models. While baseline models do not benefit from upscaling, distilled models show significant gains. Distillation closes half the gap between the baseline and teacher models. Importance-based pruning outperforms na\"\i ve pruning and even the unpruned model at larger sizes, likely due to fine-tuning. However, when too many parameters are removed, the performance of the pruned model falls below that of the distilled models. These results show that pruning bridges the gap between distilled and teacher models at larger model sizes.

The third section of \cref{tab:restable} presents correlation coefficients for baseline model variant 7, its distilled counterpart, and a pruned model at 29\% of the original size. The distilled model clearly improves on the labeled-only baseline model for most of the test datasets. While the distillation approach leads to great improvements on datasets containing VoIP-typical speech degradations, it could not carry over the generalization properties of the XLS-R-based teacher model to out-of-domain text-to-speech data. Even though French synthetic speech was in the unlabeled data for distillation, the distilled model only achieves low correlation with MOS labels on the Blizzard datasets. This indicates that distillation only works well for data on which the teacher model already achieves high correlation.


\section{Conclusion}\label{sec:concl}
We investigated distillation and pruning techniques to compress an XLS-R-based speech quality assessment model into a more compact model. The proposed distillation approach using unlabeled degraded speech data significantly outperforms baselines trained only on labeled data: The distillation-based models could close approximately half of the gap to the XLS-R-based teacher speech quality assessment model. While distillation excels for smaller target model sizes, data-driven pruning is more suitable for larger model sizes. Combining data-driven pruning and the proposed distillation technique, we achieve scalable speech quality assessment over more than three orders of magnitude of model size.


\pagebreak
\bibliographystyle{IEEEtran}
\bibliography{refs}


\end{document}